\RequirePackage{silence}
\documentclass[12pt]{article}
\usepackage[top=2.0cm,bottom=2.0cm,left=2.54cm,right=2.54cm]{geometry}
\usepackage{mathrsfs,amsmath,amsfonts,mathtools,amssymb,wasysym,graphicx,subfigure,caption,rotating,comment,booktabs,cite}
\usepackage{bm}
\usepackage{color}  
\usepackage{tikz}
\usepackage{xspace}

\usetikzlibrary{cd}

\usepackage[hyperindex=true,
          pdfstartview=FitH,
          bookmarksnumbered=true,
          bookmarksopen=true,
          citecolor=blue,
          linkcolor=blue,
          colorlinks=true,
          pdfborder=001,
          unicode]{hyperref}

\allowdisplaybreaks
\newcommand{\D}{{\rm d}}

\renewcommand\({\left(}
\renewcommand\){\right)}

\newcommand{\ind}[2]{^{#1}{}_{#2}}

\newcommand{\add}{\text{add}}
\newcommand{\fa}{\mathfrak{a}}
\newcommand{\fb}{\mathfrak{b}}
\newcommand{\LEt}{LE$_t$\xspace}
\newcommand{\LEtau}{LE$_\tau$\xspace}
\newcommand{\re}{\mathrm e}

\begin{document}
\title{Relativistic stochastic mechanics II:\\ 
Reduced Fokker-Planck equation in curved spacetime}

\author{
Yifan Cai\thanks{{\em email}: \href{mailto:caiyifan@mail.nankai.edu.cn}
{caiyifan@mail.nankai.edu.cn}},~
Tao Wang\thanks{
{\em email}: \href{mailto:taowang@mail.nankai.edu.cn}{taowang@mail.nankai.edu.cn}},~
and Liu Zhao\thanks{Corresponding author, {\em email}: 
\href{mailto:lzhao@nankai.edu.cn}{lzhao@nankai.edu.cn}}\\
School of Physics, Nankai University, Tianjin 300071, China}

\date{}
\maketitle

\begin{abstract}
The general covariant Fokker-Planck equations associated with the two different versions 
of covariant Langevin equation in Part I of this series of work are derived, 
both lead to the same reduced Fokker-Planck equation for the non-normalized
one particle distribution function (1PDF). The relationship between various distribution 
functions is clarified in this process. Several macroscopic quantities are introduced
by use of the 1PDF, and the results indicate an intimate
connection with the description in relativistic kinetic theory. The concept of 
relativistic equilibrium state of the heat reservoir is also clarified, and, 
under the working assumption that the Brownian particle should approach the same
equilibrium distribution as the heat reservoir in the long time limit, a 
general covariant version of Einstein relation arises.
\end{abstract}

\section{Introduction}
\label{Introduction}

This is Part II of our series of works on relativistic stochastic mechanics. 
Part I of this series has already been presented in \cite{cai2023relativistic}. 
The major subject of concern in Part I is the construction of manifestly general covariant
Langevin equation from the observer's perspective. Two different versions 
of relativistic Langevin equation (denoted \LEtau and \LEt respectively) were proposed,
among which \LEtau takes the proper time $\tau$ of the 
Brownian particle as evolution parameter and \LEt takes the proper time $t$ of some prescribed 
observer as evolution parameter. It was shown that although \LEtau contains some conceptual
issues from the point of view of the prescribed observer, there is numerical evidence 
indicating that \LEtau and \LEt can produce the same distribution in the same 
space of micro states (SoMS) for the case of $(1+1)$-dimensional Minkowski spacetime, 
which in turn suggests that we may be able to extract useful probability distributions 
from \LEtau.

Besides Langevin equation, Fokker-Planck equation (FPE) is another important 
equation in stochastic mechanics. The route leading from Langevin equation to 
FPE can be regarded as a bridge from mechanics to statistical 
physics. The study of FPE was initiated about a hundred years ago
\cite{fokker1914mittlere,planck1917satz}, and the purpose is to analyze the 
diffusion phenomena (in the configuration space) of suspended particles in solution. 
Kolmogoroff \cite{kolmogoroff1931analytischen} gave an explanation of the 
equation of the same form from the perspective of stochastic processes, therefore 
the corresponding equation is also called Kolmogoroff equation. 
Later, Klein \cite{klein1921statistischen} and Kramers \cite{kramers1940brownian} 
generalized the equation to the phase space. Chandrasekhar provided a detailed report 
on the relevant topics \cite{chandrasekhar1943stochastic}, and the solution 
to the Klein-Kramers equation describing a relaxation process was also given. 
All these works used the transition probability to study the evolution of 
random variables.  With the development of stochastic differential equations, 
related topics have been extensively studied by use of Ito calculus 
\cite{ito1951stochastic}, and some more modern methods 
about this topic can be found in \cite{oksendal2003stochastic}. 

In the relativistic regime, there is no Markov process satisfying causality 
on the spacetime manifold \cite{dudley1966lorentz,hakim1968relativistic}. The only 
choice is to study FPE on the SoMS ---  a subspace of the future mass shell bundle. 
This means that the equation to be considered needs to be of the Klein-Kramers type. 
However, with the usual abuse of terminology, we still use the name FPE for convenience.

The study of relativistic stochastic process can be traced back 
to Dudley \cite{dudley1966lorentz,dudley1967note} and Hakim 
\cite{hakim1965covariant,hakim1968relativistic}, who first discussed the space of 
states for stochastic processes in a model independent way. The study of concrete 
relativistic stochastic processes, e.g. the relativistic Ornstein-Uhlenbeck process, 
was carried out by Debbasch {\em et al} in \cite{debbasch1997relativistic}. 
Barbachoux {\em et al} \cite{barbachoux2001spatially,barbachoux2001covariant} made
some discussions about the corresponding FPE (Kolmogoroff equation). 
Dunkel {\em et al}\cite{dunkel2005theory1,dunkel2005theory2,dunkel2006relativistic, 
dunkel2007one,dunkel2009relativistic} also studied similar topics in the special 
relativistic context, and their model gave an intuitive understanding of the 
relativistic Brownian motion. Herrmann \cite{herrmann2010diffusion} and Haba 
\cite{haba2010relativistic} extended the studies to general relativistic context, 
with some emphasis placed on the manifest general covariance. 

It is necessary to point out that, in all previous works, the important role
played by the observer has not beed sufficiently addressed. In this work, 
we shall show that properly addressing the role played by the observer is the 
starting point in understanding different versions of general covariant 
FPE that arise either directly or indirectly from the Langevin 
equations \LEtau and/or \LEt proposed in \cite{cai2023relativistic}. 
In particular, the observer plays an important role in the interpretation of 
various distribution functions that appear in different versions of 
covariant FPE. 

Another important aspect which has not been made sufficiently clear in previous works
is the state of the heat reservoir. The description for the 
non-relativistic Brownian motion of a heavy particle inside 
a heat reservoir relies on two basic assumptions. First, the heat reservoir should 
have reached thermodynamic equilibrium, and the only impact of the reservoir 
on the Brownian particle is provided through thermal fluctuations, of which 
the fast and slow parts manifest respectively in the Langevin equation in the 
form of stochastic and damping forces. Second, the stochastic motion of 
the Brownian particle should be able to mimic a relaxation process, which means that, 
after sufficiently long time, the probability distribution for the Brownian particle 
should approach the same equilibrium distribution obeyed by the particles from 
the heat reservoir. We shall see in Sec.~\ref{local-thermal-equilibrium-einstein-relation}
that the concept of equilibrium state for the heat reservoir needs to be 
re-examined carefully in the relativistic context.

This paper is organized as follows. In Sec.~\ref{the-state-space}, 
we presents an introduction to the SoMS for the Brownian particle and prepare the 
notations to be used in the forthcoming sections. The description for the SoMS  
is placed on the same ground as in relativistic kinetic theory 
\cite{sarbach2013rel,sarbach2014geo,sarbach2014tangent}, with the expectation that 
the deep connection between these complementary approaches to non-equilibrium 
statistical physics could be further elucidated. Such a treatment is  
more appealing than the alternative approaches, e.g. those making use of jet bundles.
In Sec.~\ref{Covariant-Fokker-Planck-equations}, we 
deduce the covariant FPEs from the Langevin equations with 
different evolution parameters introduced in \cite{cai2023relativistic}. 
Sec.~\ref{reduced} is devoted to clarifying the relationship between different 
distribution functions. In this section, we also introduce a new distribution,
which is identified to be the one particle distribution function (1PDF)
in the sense of relativistic kinetic theory, together with 
its evolution equation, i.e. the reduced FPE. 
Sec.~\ref{probability-current-and-energy-current} introduces some thermodynamic 
quantities and thermodynamic relations, and the formulation seems to indicate some 
deep connections between the approaches of stochastic mechanics and relativistic 
kinetic theory. In Sec.~\ref{local-thermal-equilibrium-einstein-relation}, we clarify the 
meaning of the equilibrium state of the heat reservoir, and, by assuming that the 
1PDF should approach the intrinsic equilibrium distribution
of the heat reservoir, we deduce a general relativistic version of the Einstein 
relation. Finally, in Sec.~\ref{conclusion}, we present a brief summary of the results.

\section{The SoMS and its geometry}
\label{the-state-space}

Since this work is a followup to Ref.\cite{cai2023relativistic}, we use exactly the 
same notations and conventions as in \cite{cai2023relativistic}. In particular, 
the spacetime manifold $\mathcal M$ is taken to be a curved 
pseudo-Riemannian manifold of dimension $(d+1)$ with a mostly positive signature. 
The future mass shell bundle $\Gamma^{+}_{m}$ over $\mathcal M$ is defined as
\begin{align}\label{def-mass-shell-bundle}
\Gamma^{+}_{m}:=\{(x,p)\in T\mathcal M~|~ g_{\mu\nu}(x) p^{\mu} p^{\nu} =-m^{2}\ 
\text{and} \ p^{\mu} Z_{\mu}(x)<0\},
\end{align}
in which $Z^\mu(x)$ denotes the proper velocity of some observer field. Later on, we shall 
omit the word ``future'' and simply refer to $\Gamma^{+}_{m}$ as the mass shell bundle. 
The momentum space of the Brownian particle at the event $x$ is identified as the 
intersection of the tangent space $T_{x}\mathcal M$ with the mass shell bundle
and is referred to simply as the mass shell at $x$, 
\begin{align}\label{def-mass-shell}
\(\Gamma^{+}_{m}\)_{x}:=T_{x}\mathcal M\cap\Gamma^{+}_{m},
\end{align} 
and the configuration space is labeled by the proper time $t$ of a single prescribed
observer, {\em Alice}, as the level set
\[
\mathcal S_t:=\{x\in {\mathcal M}| t(x)=t={\rm const}.\},
\]
where $t(x)$ is an extension of the proper time $t$ over $\mathcal M$ as a scalar field. 
Denoting the proper velocity of Alice also by $Z^\mu$ should produce no confusions.
The SoMS of the Brownian particle is then given by
\begin{align}
\Sigma_t:=\bigcup_{x\in\mathcal S_t} (\Gamma^+_m)_x=\{(x,p)\in\Gamma^+_m|x\in\mathcal S_t\},
\end{align}
which is clearly observer-dependent. The above specification for the SoMS of the 
Brownian particle naturally falls inline with the tangent bundle formalism of relativistic 
kinetic theory \cite{sarbach2013rel,sarbach2014geo,sarbach2014tangent}. This will 
certainly benefit for the communication between the two important branches of non-equilibrium
relativistic statistical physics. An immediate benefit is to adopt the Sasaki metric 
\cite{sasaki1958differential} for describing the local geometry of the tangent bundle 
(and subspaces thereof). 

Before proceeding, let us introduce our conventions on indices. 
We use both concrete and abstract index notations, however with omissions of 
the abstract indices when there no confusions could arise. 
Lower-case Greek letters $\alpha,\beta,\mu,\nu,\rho,...$ denote 
concrete indices and range from $0$ to $d$. Latin capital letters $A,B,...$ 
and some lower-case Latin letters, such as $i,j,...$, also denote concrete 
indices. The upper-case Latin indices range from $0$ to $2d$ and is associated 
with tensors on the mass shell bundle, while the lower-case Latin indices $i,j,...$
range from $1$ to $d$. The other lower-case Latin letters $a,b,...$ denote 
abstract indices. Additionally, we use the calligraphy fonts, like $\mathcal F$, 
$\mathcal R$ and $\mathcal K$, to denote tensors on the momentum space $(\Gamma_m^+)_x$, 
and the cursive fonts, like $\mathscr N$, $\mathscr Z$ and $\mathscr L$, to denote 
tensors on the  mass shell bundle $\Gamma_m^+$.

Since $\Gamma^+_m$, $(\Gamma^+_m)_x$ and $\Sigma_t$ are all subspaces of the
tangent bundle $T\mathcal M$, it is appropriate to begin by describing the 
relevant geometric structures on $T\mathcal M$. What really matters is the
tangent space of the tangent bundle, which can be subdivided into the direct sum 
of horizontal and vertical subspaces \cite{dombrowski1962geometry,gudmundsson2002geometry},
\begin{align}\label{decompose-TM}
T_{(x,p)}(T\mathcal M)=H_{(x,p)}\oplus V_{(x,p)},
\end{align}
where $H_{(x,p)}$ is spanned by 
\begin{align}
e_{\mu}=\frac{\partial}{\partial x^{\mu}}-\Gamma^{\alpha}{}_{\mu\beta}
p^{\beta}\frac{\partial}{\partial p^\alpha},
\label{emu}
\end{align}
and $V_{(x,p)}$ is spanned by $\partial/\partial p^{\mu}$. Here 
$\Gamma^{\alpha}{}_{\mu\beta}$ represents the usual Christoffel
connection associated with the spacetime metric $g_{\mu\nu}$. 

The metric on the tangent bundle $T\mathcal M$ is given by the Sasaki metric, which can 
be written as a direct sum of metrics on the two subspaces,
\begin{align}\label{sasaki-metric}
\hat{g}_{ab}:=\underbrace{g_{\mu\nu}\D x^{\mu}_{\ a} 
\D x^{\nu}_{\ b}}_{\text{the metric of } H_{(x,p)}}
+\underbrace{g_{\mu\nu}\,\theta^{\mu}_{\ a}
\theta^{\nu}_{\ b}}_{\text{the metric of } V_{(x,p)}},
\end{align}
where 
\begin{align}
\theta^{\mu}=\D p^{\mu}+\Gamma^{\mu}{}_{\alpha\beta}p^{\alpha} \D x^{\beta}.
\end{align}
$\{e_{\mu},\partial/\partial p^{\mu}\}$ and $\{\D x^{\mu},\theta^{\mu}\}$ are dual bases 
on the tangent and cotangent spaces of the tangent bundle respectively. 
As a hypersurface on the tangent bundle, the mass shell bundle is naturally equipped with
an induced metric
\begin{align}\label{metric-of-mass-shell-bundle}
\hat{h}_{ab}&:=\hat{g}_{ab}+\hat N_{a}\hat N_{b},\qquad 
\hat N^a:=(m)^{-1}p^\mu\(\frac{\partial}{\partial p^{\mu}}\)^a,
\end{align}
where $\hat N^a$ is the unit normal vector of 
the mass shell bundle. The metric of the mass shell bundle can also be written as the 
direct sum of the metrics of the horizontal subspace and the momentum space,
\begin{align}\label{decompose-metric-of-mass-shell-bundle}
\hat{h}_{ab}=\underbrace{g_{\mu\nu}\D x^{\mu}{}_{a} 
\D x^{\nu}{}_{b}}_{\text{the metric of } H_{(x,p)}}
+\underbrace{\Delta_{\mu\nu}(p)\theta^{\mu}{}_{a}
\theta^{\nu}{}_{b}}_{\text{the metric of } T_{(x,p)}(\Gamma^{+}_{m})_{x}},
\end{align}
where
\[
\Delta_{\mu\nu}(p):=g_{\mu\nu}+m^{-2}p_\mu p_\nu
\]
is the orthogonal projection tensor associated with $p^\mu$. 
The inverse of the metric $\hat{h}$ reads
\begin{align}\label{inverse-bundle-metric}
\hat{h}^{ab}=g^{\mu\nu}e_{\mu}{}^{a}e_{\nu}{}^{b}
+\Delta^{\mu\nu}(p)\(\dfrac{\partial}{\partial p^{\mu}}\)^{a}
\(\dfrac{\partial}{\partial p^{\nu}}\)^{b}.
\end{align}
It is obvious that the metric on the momentum space $(\Gamma_m^+)_x$ and its inverse are
respectively
\begin{align}
h_{ab}=\Delta_{\mu\nu}(p)\theta^{\mu}{}_{a}\theta^{\nu}{}_{b},
\qquad
h^{ab}=\Delta^{\mu\nu}(p)\(\dfrac{\partial}{\partial p^{\mu}}\)^{a}
\(\dfrac{\partial}{\partial p^{\nu}}\)^{b}.
\end{align}
Please remember that we use $\hat{h}_{ab}$ for the metric on the mass shell bundle
and $h_{ab}$ for the metric on the fiber space alone.

Since $\Sigma_t$ is a hypersurface on the mass shell bundle, 
there is a normal vector field. This normal vector field is given by
\begin{align}
\mathscr{Z}^a=Z^\mu e_\mu{}^a.
\end{align}
$\mathscr{Z}^a$ is actually an up-lift of the observer field onto the 
mass shell bundle.

Using the above metrics, it is easy to find the invariant volume elements on
$T\mathcal M$, $\Gamma^{+}_{m}$ and $(\Gamma^{+}_{m})_{x}$, respectively 
\cite{sarbach2014geo},
\begin{align}
\eta_{T\mathcal M}&=g \,\D x^{0}\wedge\D x^{1}\wedge...\wedge\D x^{d}
\wedge\D p^{0}\wedge...\wedge \D p^{d}\label{volume-TM},\\
\eta_{\Gamma^{+}_{m}}&=\frac{g}{p_{0}}\D x^{0}\wedge\D x^{1}
\wedge...\wedge\D x^{d}\wedge\D p^{1}\wedge...\wedge \D p^{d}
\label{volume-mass-shell-bundle},\\
\eta_{(\Gamma^{+}_{m})_{x}}&=\frac{\sqrt{g}}{p_{0}} 
\D p^{1}\wedge...\wedge \D p^{d},
\label{volume-mass-shell}
\end{align}
where we have introduced $g=|\det(g_{\mu\nu})|$.

As mentioned above, $\{e_{\mu},\partial/\partial p^{\mu}\}$ is the basis of 
$T_{(x,p)}(T\mathcal M)$, so an arbitrary tangent vector on $T\mathcal M$ can be 
written as
\begin{align}\label{arbitrary-vector-TM}
\mathscr{V}^{a}=V^{\mu}e_{\mu}^{\ a}+\mathcal{V}^{\mu}
\(\dfrac{\partial}{\partial p^{\mu}}\)^{a}.
\end{align}
The vectors with vanishing components $V^{\mu}$ can also be treated as 
tangent vectors on the tangent space,
and these will be denoted as
\begin{align}\label{arbitrary-vector-TM-fiber}
\mathcal{V}^{a}=\mathcal{V}^{\mu}\(\frac{\partial}{\partial p^\mu}\)^{a}.
\end{align}

For tangent vectors on the mass shell $(\Gamma^+_m)_x$, it is convenient to 
introduce the following vector basis,
\begin{align}\label{basis-mass-shell}
\(\dfrac{\partial}{\partial \breve{p}^{i}}\)^{a}
:=\(\dfrac{\partial}{\partial {p}^{i}}\)^{a}
-\dfrac{p_{i}}{p_{0}}\(\dfrac{\partial}{\partial {p}^{0}}\)^{a}.
\end{align}
Notice that, due to the mass shell condition, $(\Gamma^+_m)_x$ has one less dimension than 
$T_x \mathcal M$, and so are their respective tangent spaces. Since $(\Gamma^+_m)_x$ is a 
hypersurface in $T_x \mathcal M$ with normalized normal vector $\hat{N}^a$ given in 
eq.\eqref{metric-of-mass-shell-bundle}, any tangent vector on $(\Gamma^+_m)_x$ is
automatically a tangent vector on $T_x \mathcal M$. Therefore, we can also write the tangent
vectors on $(\Gamma^+_m)_x$  in terms of the basis $\{(\partial/\partial p^{\mu})^a\}$. 
In other words, any tangent vector $\mathcal{V}^{a}$ on $(\Gamma^+_m)_x$ acquires two
different component representations
\[
\mathcal{V}^{a} = \mathcal{V}^{i}\(\dfrac{\partial}{\partial \breve{p}^{i}}\)^{a}
\quad \text{and}\quad 
\mathcal{V}^{a} = \mathcal{V}^\mu \(\frac{\partial}{\partial p^\mu}\)^{a}.
\]
It is straightforward to check that these two representations are equivalent,
\begin{align}\label{relation-vector-on-mass-shell}
\mathcal{V}^{i}\(\frac{\partial}{\partial \breve{p}^{i}}\)^a
=\mathcal{V}^{i}\(\frac{\partial}{\partial p^{i}}\)^{a}
-\mathcal{V}^{i}\frac{p_{i}}{p_{0}}\(\frac{\partial}{\partial p^{0}}\)^{a}
=\mathcal{V}^{i}\(\frac{\partial}{\partial p^{i}}\)^{a}
+\mathcal{V}^{0}\(\frac{\partial}{\partial p^{0}}\)^{a}
=\mathcal{V}^{\mu}\(\dfrac{\partial}{\partial p^{\mu}}\)^{a},
\end{align}
wherein we have used the orthogonal condition $\mathcal{V}^{\mu} p_{\mu}
=\mathcal{V}^{0} p_{0}+\mathcal{V}^{i}p_{i}=0$. Similarly, the inverse 
metric on the momentum space can be expressed in two different bases,
\begin{align}\label{inverse-mass-shell-metric}
h^{ab}=\Delta^{\mu\nu}(p)\(\dfrac{\partial}{\partial p^{\mu}}\)^{a}
\(\dfrac{\partial}{\partial p^{\nu}}\)^{b}
=\Delta^{ij}(p)\(\dfrac{\partial}{\partial \breve{p}^{i}}\)^{a}
\(\dfrac{\partial}{\partial \breve{p}^{j}}\)^{b}.
\end{align}

In order to describe the different versions of FPE, it is customary to
introduce the covariant derivatives on each of the relevant manifolds
using the standard conventions with the aid of the metrics introduced above.
However, this step can be skiped, because we only need the covariant divergences. 
For a vector $\mathscr F^{A}=(F^{\mu},\mathcal{F}^{i})$ on the mass shell bundle, 
the covariant divergence is simply given by 
\begin{align}\label{divergence-bundle}
\hat\nabla^{(\hat h)}_{A}\mathscr F^{A}=\frac{p_{0}}{g}\frac{\partial}{\partial x^{\mu}}
\left(\frac{g}{p_{0}}F^{\mu}\right)
+p_{0}\frac{\partial}{\partial\breve{p}^{i}}\left(\frac{1}{p_{0}}\mathcal{F}^{i}\right),
\end{align}
where $\hat\nabla^{(\hat h)}$ is the covariant derivative on the mass shell bundle. 
If $F^\mu=0$, $\mathscr F$ reduces into a vector on the momentum space, and the
above equation becomes
\begin{align}\label{divergence-relation}
\hat\nabla^{(\hat h)}_{A}\mathscr F^{A}={\nabla}^{(h)}_i\mathcal{F}^{i},
\end{align}
which is automatically the covariant divergence on the the momentum space, with 
${\nabla}^{(h)}$ being the corresponding covariant derivative.

Finally, let us make some remarks on the notations and conventions. For any vector 
field $\mathscr{V}^{a}$ and any scalar field $\Phi$, the map from $\Phi$ to
$\mathscr{V}^{a}$ is denoted as $\mathscr{V}^{a}[\Phi]$. On the contrary, the action 
of the vector field $\mathscr{V}^{a}$ on $\Phi$ is denoted as $\mathscr{V}(\Phi)$. 
It is crucial to distinguish these two notations in the following text.

\section{Covariant FPEs}
\label{Covariant-Fokker-Planck-equations}

In this section, we shall derive the FPE associated with each version of 
the Langevin equation presented in \cite{cai2023relativistic} 
and try to make sense of the corresponding probability 
distribution functions (PDFs). In practice, there are different ways to obtain FPE 
from Langevin equation \cite{risken1996fokker,jacobs2010stochastic}. 
To highlight the geometric interpretation, we will adopt the diffusion operator 
method \cite{bakry2014analysis}. A brief review of the method is presented 
in Appendix A, and the construction below will be made as 
brief as possible in order to focus on the physical interpretations.

\subsection{FPE associated with \LEtau}

The Langevin equation \LEtau is given as follows,
\begin{align}
\D \tilde x^\mu_\tau&=\frac{\tilde p^\mu_\tau}{m}\D \tau,
\label{dxvsdtau}\\
\D \tilde p^\mu_\tau 
&=\left[\mathcal{R}^{\mu}{}_{\mathfrak{a}}\circ_S\D \tilde w_\tau^\mathfrak{a}
+\mathcal{F}^\mu_{\text{add}}\D \tau\right]
+\mathcal{K}^{\mu\nu}U_\nu\D \tau
-\frac{1}{m} \Gamma^\mu{}_{\alpha\beta}\tilde p^\alpha_\tau \tilde p^\beta_\tau\D \tau,
\label{langevin-relativity}
\end{align}
and the meaning of each term is described in detail in \cite{cai2023relativistic}. 
Since the stochastic forces arise from thermal fluctuations from the heat reservoir, 
it is natural to expect that
\[
\mathcal{R}^{\mu}{}_{\mathfrak{a}}\to 0,\qquad
\mathcal{F}^\mu_{\text{add}}\to 0
\]
in the low temperature limit.

Since \LEtau preserves the mass shell condition, 
not all components of $\tilde p^\mu$ could be viewed as independent,
and it is appropriate to take only $\tilde p^i$ as independent random variables. 
One can introduce a corresponding probability distribution function (PDF) 
\begin{align}
\Phi_\tau(x^\mu,p^i):=\Pr[\tilde x^\mu_\tau=x^\mu,\tilde p^i_\tau=p^i]
\label{Phitau}
\end{align}
which describes the probability for the Brownian particle 
to appear at the position $x^\mu$ in the spacetime and meanwhile has the 
momentum $p^i$ at the proper time $\tau$ of the Brownian particle itself.
This PDF is pathological for two reasons. First, $\Phi_\tau(x^\mu,p^i)$ depends on two 
time variables $\tau$ and $x^0$, which makes it hard to assign a physical interpretation; 
Second, $\Phi_\tau(x^\mu,p^i)$ is not a distribution on the 
SoMS $\Sigma_t$, but rather on the full mass shell bundle $\Gamma^+_m$. 
However, there is no technical obstacle which prevents us from constructing the 
FPE obeyed by $\Phi_\tau(x^\mu,p^i)$.

In order to get the desired FPE, we need to construct 
the diffusion operator of eq.\eqref{langevin-relativity}. For Stratonovich type 
Langevin equation, the diffusion operator can always be written in the form
\begin{align}
\mathbf{A}=\frac{\delta^{\fa\fb}}{2}L_{\mathfrak{a}}L_{\mathfrak{b}}+L_{0}.
\end{align}
In the case of eq.\eqref{langevin-relativity}, we have 
\begin{align}
L_{\mathfrak{a}}&=\mathcal{R}^{\mu}{}_{\mathfrak{a}}\frac{\partial}{\partial p^{\mu}}
=\mathcal{R}^{i}{}_{\mathfrak{a}}\frac{\partial}{\partial \breve{p}^{i}},\\
L_{0}&=\frac{p^{\mu}}{m}\frac{\partial}{\partial x^{\mu}}
-\frac{1}{m}\varGamma^{\mu}{}_{\alpha\beta}p^{\alpha} p^{\beta}
\frac{\partial}{\partial p^{\mu}}
+\(\mathcal{F}^{\mu}_{\text{add}}+\mathcal{K}^{\mu\nu}U_{\nu}\)
\frac{\partial}{\partial p^{\mu}}\notag\\
&=\frac{1}{m}\mathscr{L}+\(\mathcal{F}^{i}_{\text{add}}+\mathcal{K}^{i\nu}U_{\nu}\)
\frac{\partial}{\partial \breve{p}^{i}},
\end{align}
where $\mathscr{L}=p^\mu e_\mu$ is the Liouville vector field \cite{sarbach2014geo}.
Using the volume element of mass shell bundle eq.\eqref{volume-mass-shell-bundle}, 
the adjoint of coordinate derivative operators can be obtained straightforwardly,
\begin{align}
\(\frac{\partial}{\partial x^{\mu}}\)^*
&=-\frac{p_{0}}{g}\frac{\partial}{\partial x^{\mu}}\frac{g}{p_{0}},
\label{adjoint-x} \\
\(\frac{\partial}{\partial \breve{p}^{i}}\)^*
&=-\frac{p_{0}}{g}\frac{\partial}{\partial \breve{p}^{i}}\frac{g}{p_{0}}
=-\frac{p_0}{\sqrt{g}}\frac{\partial}{\partial \breve{p}^{i}}
\frac{\sqrt{g}}{p_0}\label{adjoint-p}.
\end{align}
With these operators, the adjoint of $L_\mathfrak{a}$ and $L_0$ can be obtained,
which read
\begin{align}
L^{*}_{\mathfrak{a}}=-{\nabla}^{(h)}_{i}\mathcal{R}^{i}{}_{\mathfrak{a}},\qquad 
L^{*}_{0}=-\frac{1}{m}\mathscr{L}-{\nabla}^{(h)}_{i}
\(\mathcal{F}^{i}_{\text{add}}+\mathcal{K}^{i\nu}U_{\nu}\),
\end{align}
where we have used $\mathscr{L}^{*}=-\mathscr{L}$. 
The FPE can then be written as
\begin{align}
\partial_{\tau} \Phi_\tau&=\mathbf{A}^{*}\Phi_\tau\notag\\
&=\(\frac{\delta^{\fa\fb}}{2}L^{*}_{\mathfrak{a}}L^{*}_{\mathfrak{b}}
+L^{*}_{0}\)\Phi_\tau\notag\\
&=\frac{\delta^{\fa\fb}}{2}{\nabla}^{(h)}_{i}\mathcal{R}^{i}{}_{\mathfrak{a}} 
{\nabla}^{(h)}_{j}\mathcal{R}^{j}{}_{\mathfrak{b}}\Phi_\tau
-{\nabla}^{(h)}_{i}\(\mathcal{F}^{i}_{\text{add}}\Phi_\tau
+\mathcal{K}^{i\nu}U_{\nu}\Phi_\tau\)
-\frac{1}{m}\mathscr{L}(\Phi_\tau)\notag\\
&={\nabla}^{(h)}_{i}\left[\frac{1}{2}\mathcal{D}^{ij}{\nabla}^{(h)}_{j}\Phi_\tau
+\frac{\delta^{\fa\fb}}{2}\(\mathcal{R}^{i}{}_{\mathfrak{a}} 
{\nabla}^{(h)}_{j}\mathcal{R}^{j}{}_{\mathfrak{b}}\)\Phi_\tau
-\mathcal{F}^{i}_{\text{add}}\Phi_\tau-\mathcal{K}^{i\nu}U_{\nu}\Phi_\tau\right]
-\frac{1}{m}\mathscr{L}(\Phi_\tau),
\label{FPPhitau}
\end{align}
where we have introduced the diffusion tensor $\mathcal{D}^{\mu\nu}
:=\mathcal{R}^{\mu}{}_{\mathfrak{a}}\mathcal{R}^{\nu}{}_{\mathfrak{a}}$. 
Defining the vector field
\begin{align}
\mathcal I^a[\Phi_\tau]:=\left[\frac{1}{2}\mathcal{D}^{ij}{\nabla}^{(h)}_j\Phi_\tau
+\frac{\delta^{\fa\fb}}{2}\(\mathcal{R}^{i}{}_{\mathfrak{a}}
{\nabla}^{(h)}_j\mathcal{R}^j{}_\mathfrak{b}\)\Phi_\tau
-\mathcal{F}^i_{\text{add}}\Phi_\tau-\mathcal{K}^{i\nu}U_\nu\Phi_\tau\right]
\(\frac{\partial}{\partial \breve{p}^i}\)^a,
\end{align}
the FPE for $\Phi_\tau$ can be written in more concise form
\begin{align}\label{relativistic-FP}
\partial_\tau\Phi_\tau={\nabla}^{(h)}_i \mathcal I^i[\Phi_\tau]
-\frac{1}{m}\mathscr{L}(\Phi_\tau).
\end{align}
Eq.\eqref{relativistic-FP} can be viewed as a continuity equation for the 
PDF $\Phi_\tau$ and its associated probability flow $ \mathscr{J}[\Phi_\tau]$, 
which is defined as
\begin{align}\label{flow-Phi}
\mathscr{J}[\Phi_\tau]:=\frac{\Phi_\tau}{m} \mathscr{L} - \mathcal I[\Phi_\tau].
\end{align}
Here, the term proportional to the Liouville vector field corresponds to the 
contribution from the free motion of the Brownian particle, 
while $\mathcal I[\Phi_\tau]$ represents the contribution from the 
interaction between the Brownian particle and the heat reservoir.

Please be reminded that 
we use the term ``flow'' instead of ``current'' to refer to the spatial components of 
the objects which obey the continuity equation. The term ``current'' is reserved for 
the full object, including the temporal component. Using the definition \eqref{flow-Phi}, 
eq. \eqref{relativistic-FP} can be rewritten in the form 
\begin{align}\label{relativistic-FP2}
  \partial_\tau\Phi_\tau
  +\hat \nabla_A^{(\hat h)} \mathscr{J}^A[\Phi_\tau]=0.
\end{align}
Eq.\eqref{relativistic-FP2} implies that the surface integral
\begin{align}
-\int_{\Sigma} \eta_{\Sigma_t}\mathscr{Z}_{A} \mathscr{J}^{A}[\Phi_\tau] 
\label{prtausigma}
\end{align}
should be the probability that the Brownian particle passes through the subarea 
$\Sigma$ in the SoMS $\Sigma_t$ per unit proper time of the Brownian particle. 
Although it looks puzzling to understand eq.\eqref{relativistic-FP2} as a continuity
equation because of the presence of two time variables, this equation still plays a 
key role while making connection to the alternative
distribution function to be introduced shortly.

\subsection{FPE associated with \LEt}

The second Langevin equation proposed in \cite{cai2023relativistic}, i.e. \LEt, 
arises from a reparametrization of \LEtau. The concrete form of \LEt reads
\begin{align}
\D \tilde y^\mu_t&=\frac{\tilde k^\mu_t}{m}\gamma^{-1}\D t, \label{yt}\\
\D \tilde k^\mu_t 
&=\left[\hat{\mathcal{R}}^{\mu}{}_{\mathfrak{a}}\circ_S\D\tilde W_t^\mathfrak{a}
+\hat{\mathcal{F}}^\mu_{\text{add}}\D t\right]
+\hat{\mathcal{K}}^{\mu\nu}U_\nu\D t 
-\frac{1}{m} \Gamma^\mu{}_{\alpha\beta}\tilde k^\alpha_t \tilde k^\beta_t \gamma^{-1} \D t,
\label{langevin-relativity-observer}
\end{align}
where $t$ represents the proper time of Alice, the prescribed observer, and
\[
\tilde y^\mu(t) = \tilde x^\mu(\tau(t)),\qquad
\tilde k^\mu(t) = \tilde p^\mu(\tau(t)).
\]
$\hat{\mathcal R}\ind{\mu}{\fa}, \hat{\mathcal{K}}^{\mu\nu}$ 
and $\hat{\mathcal{F}}_\add^\mu$
are connected with their respective un-hatted counterparts via
\begin{align}
\hat{\mathcal R}\ind{\mu}{\fa}:=\gamma^{-1/2}\mathcal R\ind{\mu}{\fa},
\quad
\hat{\mathcal{K}}^{\mu\nu}:=\gamma^{-1}{\mathcal{K}}^{\mu\nu},
\quad
\hat{\mathcal{F}}_\add^\mu:=\gamma^{-1}\mathcal F^\mu_\add
-\frac{\delta^{\fa\fb}}{2}\mathcal R\ind{\mu}{\fa}\mathcal R\ind{j}{\fb}
(\gamma^{-1/2}\hat\nabla_j \gamma^{-1/2}),
\end{align}
and
\begin{align}
\lambda := |\nabla t|,\qquad 
\gamma(\tilde x,\tilde p):=-\frac{\lambda Z_\mu \tilde p^\mu}{m}.
\end{align}
$\gamma(\tilde x,\tilde p)$ plays the role of a local Lorentz factor, 
i.e. $\D \tau = \gamma^{-1}\D t$, which is also 
random-valued because of the random motion of the Brownian particle. 
The fact that the proper time $\tau$ of the Brownian particle becomes a random variable 
from the observer's perspective is the reason why the reparametrization leading from
\LEtau to \LEt is unavoidable.

The PDF for the Brownian particle described by 
eqs.\eqref{yt}-\eqref{langevin-relativity-observer} is
\begin{align}
\Psi_t(x^\mu,p^i):=\Pr[\tilde y_{t}^\mu=x^\mu,\tilde k_t^i=p^i].
\label{Psit}
\end{align} 
Apparently, this PDF is also a two-time distribution, just like $\Phi_\tau(x^\mu,p^i)$
given in eq.\eqref{Phitau}, which is hard to understand physically. However, 
the PDF $\Psi_t(x^\mu,p^i)$ actually encodes the physical PDF $f(x^\mu,p^i)$ on 
$\Sigma_t$ in the following manner. Recall that $\Sigma_t$ can be regarded as a 
hypersurface on the mass shell bundle with normal vector field $\mathscr{Z}^a$. 
This relationship allows us to introduce an invariant volume form on $\Sigma_t$,
i.e.
\begin{align}\label{volume-element-mirco}
(\eta_{\Sigma_t})_{a_1,...,a_{2d}}:=\mathscr{Z}^{a_0}
(\eta_{\Gamma^+_m})_{a_0,a_1,...,a_{2d}}.
\end{align}
Since $t$ is the proper time of Alice, there is no randomness in $t$, 
therefore, using the co-area formula \cite{nicolaescu2011coarea,negro2022sample} of 
geometric measure theory, we can write
\begin{align}
\Psi_t(x^\mu,p^i)=\lambda\delta(t(x)-t) f(x^\mu,p^i),
\label{def_of_f}
\end{align}
in which $f(x^\mu,p^i)$ is the desired physical PDF on $\Sigma_t$. Let us stress 
that the volume elements associated with $\Psi_t(x^\mu,p^i)$ and $f(x^\mu,p^i)$
are, respectively, $\eta_{\Gamma^+_m}$ and $\eta_{\Sigma_t}$.

Following a similar procedure which leads to the FPE 
\eqref{FPPhitau}, we can get the FPE for $\Psi_t(x^\mu,p^i)$, 
which is associated with \LEt, 
\begin{align}\label{73}
\frac{\partial}{\partial t}\Psi_t+\frac{1}{m}\mathscr{L}(\gamma^{-1}\Psi_t)
=\nabla^{(h)}_{i}\left[\frac{1}{2}\hat{\mathcal{D}}^{ij}\nabla^{(h)}_j\Psi_t
+\frac{\delta^{\fa\fb}}{2}\(\hat{\mathcal{R}}^{i}_{\mathfrak{a}} 
\nabla^{(h)}_j\hat{\mathcal{R}}^{j}_{\mathfrak{b}}\)\Psi_t
-\hat{\mathcal{F}}^{i}_{\text{add}}\Psi_t-\hat{\mathcal{K}}^{i\nu}U_{\nu}\Psi_t\right],
\end{align}
where $\hat{\mathcal D}^{\mu\nu}:=\gamma^{-1}\mathcal D^{\mu\nu}$.

Now since
\begin{align}
\left[\frac{\partial}{\partial t}+\gamma^{-1}\frac{p^\mu}{m}\frac{\partial}{\partial x^\mu}
\right]\delta(t(x)-t)&=
\left[-1+\gamma^{-1}\frac{p^\mu}{m}\frac{\partial t}{\partial x^\mu}
\right]\delta'(t(x)-t)\notag\\
&=
\left[-1+\gamma^{-1}\frac{\D t}{\D \tau}\right]\delta'(t(x)-t)=0,
\end{align}
substituting eq.\eqref{def_of_f} into the left hand side of eq.\eqref{73} yields
\begin{align}
\frac{\partial}{\partial t}\Psi_t+\frac{1}{m}\mathscr{L}(\gamma^{-1}\Psi_t)
%&=\lambda f \left[
%\frac{\partial}{\partial t}+\gamma^{-1}\frac{p^\mu}{m}\frac{\partial}{\partial x^\mu}
%\right]\delta(t(x)-t)+\delta(t(x)-t)\frac{1}{m}\mathscr{L}(\gamma^{-1}\lambda f)\notag\\
&=\delta(t(x)-t)\frac{1}{m}\mathscr{L}(\gamma^{-1}\lambda f).
\end{align}
On the other hand, the first three terms in the square bracket on the right hand side of 
eq.\eqref{73} can be rearranged in the form
\begin{align}
&\frac{1}{2}\hat{\mathcal{D}}^{ij}\nabla^{(h)}_j\Psi_t
+\frac{\delta^{\fa\fb}}{2}\(\hat{\mathcal{R}}^{i}_{\mathfrak{a}} 
\nabla^{(h)}_j\hat{\mathcal{R}}^{j}_{\mathfrak{b}}\)\Psi_t
-\hat{\mathcal{F}}^{i}_{\text{add}}\Psi_t\nonumber\\
&\qquad =\frac{1}{2}{\mathcal{D}}^{ij}\nabla^{(h)}_j(\gamma^{-1}\Psi_t)
+\frac{\delta^{\fa\fb}}{2}\(\mathcal{R}^{i}_{\mathfrak{a}} 
\nabla^{(h)}_j{\mathcal{R}}^{j}_{\mathfrak{b}}\)(\gamma^{-1}\Psi_t)
-{\mathcal{F}}^{i}_{\text{add}}(\gamma^{-1}\Psi_t).
\end{align}
Therefore, the substitution of eq.\eqref{def_of_f} into eq.\eqref{73} 
yields  
\begin{align}\label{relativistic-FP-t}
\frac{1}{m}\mathscr{L}(\gamma^{-1}\lambda f)
=\nabla^{(h)}_i\mathcal I^i[\gamma^{-1}\lambda f],
\end{align}
where $\mathcal I^i[\gamma^{-1}\lambda f]$ is defined in a similar fashion as 
in eq.\eqref{relativistic-FP}.

Notice that the FPEs \eqref{relativistic-FP} and 
\eqref{relativistic-FP-t} have a similar form. By dropping the time derivative 
term $\partial_\tau\Phi_\tau$ in eq.\eqref{relativistic-FP} and 
replacing $\Phi_\tau$ with $\gamma^{-1}\lambda f$, eq.\eqref{relativistic-FP} 
can be changed into eq.\eqref{relativistic-FP-t}. This is certainly not a 
coincidence, and we will demonstrate in the next section how 
eq.\eqref{relativistic-FP} is intimately related to eq.\eqref{relativistic-FP-t}.

\section{Reduced FPE}
\label{reduced}

In Part I of this series of research \cite{cai2023relativistic}, we used numerical method 
to investigate whether the random paths generated by \LEtau and \LEt produce 
the same physical PDF on the SoMS $\Sigma_t$. The results
in the example case of $(1+1)$-dimensional Minkowski spacetime
indicate that nearly identical distributions arise from the two Langevin equations
\LEtau and \LEt. In this section, we will provide an analytical proof in generic 
spacetimes. During this proof, we will introduce a new distribution function, 
$\varphi$, together with its evolution equation, which we call the reduced 
FPE. 

Recall from eq.\eqref{prtausigma} that the integral $\displaystyle -\int_{\Sigma} 
\eta_{\Sigma_t}\mathscr{Z}_{A} \mathscr{J}^{A}[\Phi_\tau]$ represents the 
probability that the Brownian particle passes through the subregion 
$\Sigma$ in the SoMS $\Sigma_t$ per unit proper time of the Brownian particle.
From the observer's perspective, the condition ``per unit proper time of the 
Brownian particle'' is irrelevant, the actual probability that the Brownian 
particle passes through the subarea $\Sigma$ should read
\begin{align}\label{The_particle_is_through}
&\Pr[\text{The particle passes through } \Sigma]
=-\int_\mathbb{R} \D\tau\int_{\Sigma} \eta_{\Sigma_t}\mathscr{Z}_A 
\mathscr{J}^A[\Phi_\tau]\nonumber\\
&\qquad =-\int_{\Sigma} \eta_{\Sigma_t}\mathscr{Z}_A \mathscr{J}^A[\varphi],
\end{align}
where we have introduced
\begin{align}
\varphi(x,p):=\int_\mathbb{R}\D \tau \,\Phi_\tau(x,p).
\end{align}
Since $\Sigma$ is an arbitrary subregion in the SoMS $\Sigma_t$, the integrand 
$-\mathscr Z_A\mathscr J^A[\varphi]$ in eq.\eqref{The_particle_is_through} 
should be the PDF for the intersection points of the random paths with the 
SoMS $\Sigma_t$, i.e.
\begin{align}\label{prob-massshell}
{f}=-\mathscr{Z}_{A}\mathscr{J}^{A}[\varphi]
=-\dfrac{1}{m}Z_{\mu}p^{\mu}\varphi=\gamma\lambda^{-1} \varphi.
\end{align}

Now let us consider a scenario in which the random paths of the Brownian particles are 
infinitely stretched, i.e. extending from $\tau=-\infty$ to $\tau=\infty$. 
It is natural to introduce the boundary conditions
\begin{align}
\Phi_{-\infty}(x,p)=\Phi_{+\infty}(x,p)=0
\end{align}
for the PDF $\Phi_\tau$, because otherwise $\Phi_\tau$ will not be normalizable. 
Then, by integrating eq.\eqref{relativistic-FP} with respect to $\tau$, 
we can get
\begin{align}\label{reduced-relativsitic-FP}
\frac{1}{m}\mathscr{L}(\varphi)=\nabla^{(h)}_{i}\mathcal I^i[\varphi],
\end{align}
where again $\mathcal I^i[\varphi]$ is defined in a similar way as 
in eq.\eqref{relativistic-FP}.

Since $\varphi$ differs from the true PDF $f$ by the scalar factor $\gamma\lambda^{-1}$, 
it cannot be a normalized PDF. Therefore, the equation \eqref{reduced-relativsitic-FP}
obeyed by $\varphi$ will not be referred to as FPE, but
rather as {\em reduced FPE}. Bearing in mind the relationship
\eqref{prob-massshell}, one can easily recognize that eq.\eqref{reduced-relativsitic-FP}
is actually identical to eq.\eqref{relativistic-FP-t}. In other words, both the 
FPEs \eqref{relativistic-FP} and \eqref{73} give rise to the same 
reduced FPE. This fact gives an analytical evidence for the 
correctness of the numerical test presented in \cite{cai2023relativistic}.

Some remarks are in due here.

(1) Since the reduced FPE \eqref{reduced-relativsitic-FP}
is homogeneous in $\varphi$ and there is no need to normalize $\varphi$, 
there is a freedom to multiply $\varphi$ with a constant factor which preserves 
eq.\eqref{reduced-relativsitic-FP}. This freedom will be used in the next section 
while defining the particle number density of the Brownian particle.

(2) There is a common misconception about the role of $\Phi_\tau$ 
regarding a particular case, i.e. the stationary distribution $\Phi_\tau(x,p):=\Phi(x,p)$,
which is often considered to be identical to the equilibrium distribution 
for the particles of the heat reservoir, i.e. the J\"uttner distribution. 
Technically it is true that, when $\Phi_\tau$ is independent of $\tau$, eq.
\eqref{relativistic-FP} will take the same form as eq.\eqref{reduced-relativsitic-FP}. 
However, this coincidence does not imply that $\varphi(x,p)$ is identical to the 
stationary distribution $\Phi(x,p)$. There are two primary reasons for this
difference: (i) The stationary distribution is a distribution which does not change 
with the time of some stationary observer, rather than of the Brownian particle; 
(ii) The identification of $\varphi(x,p)$ with the stationary distribution $\Phi(x,p)$ 
implies that the reduced FPE can only describe the stationary 
states, whereas it can actually describe the whole relaxation process, 
as will be demonstrated in Sec.~\ref{probability-current-and-energy-current}
and Sec.~\ref{local-thermal-equilibrium-einstein-relation}.

Following a similar fashion with eq.\eqref{flow-Phi}, we can introduce a {\em current}
associated with $\varphi$,
\[
\mathscr{J}[\varphi]:=\frac{\varphi}{m} \mathscr{L} - \mathcal I[\varphi].
\]
Then the reduced FPE \eqref{reduced-relativsitic-FP} could be rewritten 
as the current conservation equation 
\begin{align}\label{current-conservation-mass-shell}
\hat\nabla^{(\hat h)}_{A} \mathscr{J}^{A}[\varphi]=0
\end{align}
on the mass shell bundle. Let us stress that $\mathscr{J}[\varphi]$ is now interpreted as 
a {\em current}, rather than {\em flow}, because only a single time variable is present 
in the above equation which is hidden behind the index $A$.
The conservation of the current $\mathscr{J}^{A}[\varphi]$ does not 
correspond to conservation of probability, but rather to conservation of matter. 
More details on this point will be presented in the next section.

\section{Macroscopic quantities and interpretation of $\varphi(x,p)$}
\label{probability-current-and-energy-current}

Let $\mathcal S$ be an arbitrary subregion in the configuration space $\mathcal S_t$, 
and $\Sigma:=\{(x,p)\in\Gamma^+_m|x\in\mathcal{S}\}$ is the corresponding subregion
in the SoMS. When Alice is not bound together with the coordinate system, 
the proper time $t$ will be different from the coordinate time $x^0$, 
which means that $\mathcal S_t$ is not the coordinate hypersurface with fixed $x^0$, 
but rather a tilted hypersurface with mixtures between $x^0$ and $x^i$. 
Nevertheless, since the PDF $f(x,p)$ is by definition the probability density on 
$\Sigma_t$, and that $\eta_{\Sigma_t}=\eta_{\mathcal{S}_t}\wedge\eta_{(\Gamma^+_m)_x}$,
we can calculate the probability for the Brownian particle to appear in $\mathcal S$
at the time $t$ as
\begin{align}
\Pr[\mathcal{S}]&=\Pr[\Sigma]=\int_{\Sigma}\eta_{\Sigma_t}f
\nonumber\\
&=-\int_{\Sigma}\eta_{\Sigma_t}\mathscr{Z}_{a}\mathscr{J}^{a}[\varphi]
=-\int_{\mathcal{S}}\eta_{\mathcal{S}_{t}}Z_{\mu}
\int_{(\Gamma_{m}^{+})_{x}}\eta_{(\Gamma_{m}^{+})_{x}}\dfrac{p^{\mu}}{m}\varphi.
\label{partialprob}
\end{align}
The change from $f$ to $\varphi$ in the integrand of the last equality reflects the 
tiltedness of $\mathcal S_t$ in the spacetime. 

Now consider the case with $N$ non-interacting Brownian particles coexisting in the same
heat reservoir. By putting an extra factor $N$ in front of the integrals in 
eq.\eqref{partialprob} and enlarging $\Sigma$ into $\Sigma_t$, we should get 
$N$ as the final value of the integration. Therefore, by dropping the integral
over $\mathcal S$, we get the particle number density in the configuration space
\begin{align}
\bar n=- N Z_{\mu}\int_{(\Gamma_{m}^{+})_{x}}\eta_{(\Gamma_{m}^{+})_{x}}
\dfrac{p^{\mu}}{m}\varphi.
\end{align}
Recall that the particle number density should be defined as 
\[
\bar n= - Z_\mu N^\mu[\varphi],
\]
wherein $N^\mu$ denotes the particle number current. At present, 
the particle number current reads
\begin{align}\label{defofprobabilitycurrentonspacetime}
N^{\mu}[\varphi]= \int_{(\Gamma_{m}^{+})_{x}}
\eta_{(\Gamma_{m}^{+})_{x}}\dfrac{p^{\mu}}{m}N \varphi.
%:=\int_{(\Gamma_{m}^{+})_{x}}
%\eta_{(\Gamma_{m}^{+})_{x}}\dfrac{p^{\mu}}{m}\phi.
\end{align}
It is remarkable that the above form of the particle number current is 
identical to that given in relativistic kinetic theory 
(except for the constant factor $N$), provided that 
$\varphi$ is identified with the 1PDF which obeys 
the relativistic Boltzmann equation. This resemblance reminds us that there
may be some deep connections between the approaches of relativistic stochastic mechanics
and of relativistic kinetic theory. 

Since there is no chemical reactions between the Brownian particles, 
the particle current must be conserved. This fact can be proved 
using Stokes' theorem. Let $V$ be an region in the spacetime manifold $\mathcal M$, 
and $\Gamma=\{(x,p)\in\Gamma^{+}_{m}|\ x\in V \}$ is the corresponding region on 
the mass shell bundle. Let $n^{\mu}$ be the unit normal vector field of 
$\partial V$ which induces the unit normal vector field $\mathscr{N}$ of 
$\partial \Gamma$. The Stokes' theorem on the mass shell bundle reads
\begin{align}
\int_\Gamma \eta_{\Gamma^+_m} \hat\nabla^{(\hat h)}_{A} \mathscr{J}^A[\varphi]
=\int_{\partial\Gamma} \eta_{\partial\Gamma} \mathscr{N}_A \mathscr{J}^A[\varphi]. 
\end{align}
Using the fact that $\partial\Gamma=\{(x,p)\in \Gamma^{+}_{m}|\ x\in\partial V \}$
and that $\mathscr{N}^a=n^\mu e_\mu{}^a$, we can rewrite the above equation as
\begin{align}
&\int_V \eta_{\mathcal M} \int_{(\Gamma^+_m)_x}\eta_{(\Gamma^+_m)_x} 
\hat\nabla^{(\hat h)}_{A}\mathscr{J}^A[\varphi]
=\int_{\partial V}\eta_{\partial V} \int_{(\Gamma^+_m)_x}\eta_{(\Gamma^+_m)_x}
\mathscr{N}_A \mathscr{J}^A[\varphi]\nonumber\\
&\qquad
=\frac{1}{N}\int_{\partial V}\eta_{\partial V} n_\mu N^\mu[\varphi]
=\frac{1}{N}\int_V \eta_{\mathcal M} \nabla_\mu N^\mu[\varphi],
\end{align}
where $\nabla_\mu$ denotes the usual covariant derivative on the spacetime manifold.
Due to the arbitrariness of $V$, we can drop the integration with respect to the measure 
$\eta_{\mathcal M}$ and get
\begin{align}\label{current-conservation-spacetime}
\nabla_\mu N^\mu[\varphi]
=N \int_{(\Gamma^+_m)_x}\eta_{(\Gamma^+_m)_x}
\hat\nabla^{(\hat h)}_{A} \mathscr{J}^A[\varphi]=0,
\end{align}
which means that $N^{\mu}[\varphi]$ is a conservation current on the spacetime.

The energy of a single Brownian particle measured by Alice 
is defined as
\begin{align}
E:=-p^\mu Z_\mu.
\end{align}
Thus the single particle contribution to the average energy flux through the 
subregion $\Sigma$ in the SoMS $\Sigma_t$ should read
\begin{align}\label{ensemble-energy}
\bar{E}[\Sigma]&:=\int_\Sigma \eta_{\Sigma} 
\mathscr Z_{A} \mathscr{J}^{A}[\varphi]p^{\mu} Z_{\mu} 
=\int_{\mathcal{S}} \eta_{\mathcal{S}} Z_{\mu} Z_{\nu} 
\int_{(\Gamma^+_{m})_{x}}\eta_{(\Gamma^{+}_{m})_{x}}\frac{p^{\nu} p^{\mu}}{m}\varphi.
\end{align}
The second integration factor, i.e.
\begin{align}\label{energy-momentum}
T^{\mu\nu}[\varphi]:=\int_{(\Gamma^{+}_{m})_{x}}\eta_{(\Gamma^{+}_{m})_{x}}
\frac{p^{\nu} p^{\mu}}{m}\varphi,
\end{align}
is recognized to be the single particle contribution to the energy-momentum tensor,
and 
\begin{align}
\rho:= Z_{\mu} Z_{\nu} T^{\mu\nu}[\varphi]
= Z_{\mu} Z_{\nu} 
\int_{(\Gamma^+_{m})_{x}}\eta_{(\Gamma^{+}_{m})_{x}}\frac{p^{\nu} p^{\mu}}{m}\varphi
\end{align} 
is naturally the single particle contribution to the energy density.

The single particle contribution to the average energy-momentum vector 
of the Brownian particle is defined as
\[
E^\mu[\varphi]:=-Z_\nu T^{\mu\nu}[\varphi].
\]
In general, $E^\mu[\varphi]$ is non-conserved because of the joint effects of 
gravitational work and heat transfer from the heat reservoir. 
Since
\begin{align}
&- \int_V \eta_{\mathcal{M}} \nabla_{\mu} (Z_{\nu} T^{\mu \nu}[\varphi])
= - \int_{\partial V} \eta_{\partial V} n_{\mu} Z_{\nu} T^{\mu \nu}[\varphi]
= -\int_{\partial V} \eta_{\partial V} \int_{(\Gamma^+_m)_x}
\eta_{(\Gamma^+_m)_x} 
(Z_\nu p^\nu) n_{\mu} \frac{p^{\mu}}{m} \varphi\nonumber\\
&\qquad =\int_{\partial V} \eta_{\partial V} 
\int_{(\Gamma^+_m)_x} \eta_{(\Gamma^+_m)_x}  \mathscr{N}_A \mathscr{J}^A[\varphi] E
=\int_V \eta_{\mathcal{M}} \int_{(\Gamma^+_m)_x} \eta_{(\Gamma^+_m)_x}
\hat\nabla^{(\hat h)}_A (E \mathscr{J}^A [\varphi]),
\end{align}
we have
\begin{align}
\nabla_\mu E^\mu[\varphi] 
&= -\nabla_\mu \(Z_\nu T^{\mu\nu}[\varphi]\)
=-\int_{(\Gamma^+_m)_x}\eta_{(\Gamma^+_m)_x}
\hat\nabla^{(\hat h)}_A \(p^\nu Z_\nu \mathscr{J}^A[\varphi]\)\notag\\
&=-\int_{(\Gamma^+_m)_x}\eta_{(\Gamma^+_m)_x}\mathscr{J}^A[\varphi]
\hat\nabla^{(\hat h)}_A \(p^\nu Z_\nu \)\notag\\
&=-\int_{(\Gamma^+_m)_x}\eta_{(\Gamma^+_m)_x}\(\frac{\varphi}{m}\mathscr{L}(Z_\nu p^\nu)
-Z_\nu\mathcal I^\nu[\varphi]\)\notag\\
&=-T^{\mu\nu}[\varphi]\nabla_\mu Z_\nu
+Z_\nu\int_{(\Gamma^+_m)_x}\eta_{(\Gamma^+_m)_x}\mathcal I^\nu[\varphi].
\label{nabE}
\end{align}
The first term on the right hand side of eq.\eqref{nabE} is the average 
gravitational power acting on the Brownian particle, i.e.
\begin{align}
P_{\text{grav}}[\varphi]&:=-T^{\mu\nu}[\varphi]\nabla_\mu Z_\nu 
= \int_{(\Gamma^{+}_{m})_{x}}\eta_{(\Gamma^{+}_{m})_{x}} 
\mathcal P_{\text{grav}}(Z) \varphi,
\end{align}
where
\begin{align}
\mathcal P_{\text{grav}}(Z)=-\frac{p^\mu p^\nu}{m}\nabla_\mu Z_\nu
\end{align}
is the the gravitational power along a single trajectory of the particle 
\cite{liu2020work} as measured by Alice. Thus the second term on the right hand side of 
eq.\eqref{nabE} should be interpreted as the heat transfer rate from the heat reservoir,  
\begin{align}
Q[\varphi]:=\int_{(\Gamma^+_m)_x}\eta_{(\Gamma^+_m)_x} Z_\nu \mathcal I^\nu[\varphi]
=-Z_\nu \nabla_\mu T^{\mu\nu}[\varphi].
\label{heat-current}
\end{align}
In the end, we have
\begin{align}
\nabla_\mu E^\mu[\varphi]=P_{\text{grav}}[\varphi]+Q[\varphi],
\end{align}
which is reminiscent to the first law of thermodynamics, but is presented in terms 
of the divergence of the average energy-momentum vector, the gravitational power 
and the heat transfer rate. Please note that the last equation is valid for any 
observer. However, for different observers, 
the values of $P_{\text{grav}}[\varphi]$ and $Q[\varphi]$ can be different.

\section{Relativistic equilibrium state and Einstein relation}
\label{local-thermal-equilibrium-einstein-relation}

So far, we have not paid a word on the state of the heat reservoir, except for the implicit
assumption of an equilibrium state. This does not make any harm to the formal construction
of FPE. However, when the solution to the FPE
is concerned, an explicit description for the equilibrium state of the reservoir 
becomes inevitable.

As mentioned in the introduction, there are two basic assumptions in the description 
for the Brownian motion of a heavy particle in a heat reservoir.
In the relativistic context, these assumptions need to be re-examined. 

The first problem one encounters is the proper definition for the equilibrium state of the 
reservoir. It is well known that, in the presence of gravity, a macroscopic  
system cannot reach the thermodynamic equilibrium in the usual sense, i.e. the one with 
uniform temperature and chemical potential. The reason lies in that there is a 
bilateral interaction between thermal and gravitational effects. On the one hand, 
thermal energy as a form of energy should generate gravity; on the other hand, 
gravity, as a long range interaction, has nontrivial impact on the relaxation process, 
leading to the final state with non-uniform temperature and 
chemical potential. 

Meanwhile, the choice of observer also brings about some
subtleties in describing the state of the heat reservoir. 
The importance of the role of observer can be revealed in two different 
aspects: i) According to the equivalence principle, gravity is 
locally indistinguishable from acceleration. Therefore, the strength of the  
gravitational force experienced by the observer and by the macroscopic system 
being observed could be different, provided the amounts of 
accelerations are different. ii) There has long been a dispute on the 
of relativistic transformation rules of thermodynamic parameters, mostly about 
the transform of temperature, but also include the transform of chemical potential. 
According to the results of \cite{hao2022relativistic}, these transformation rules 
are related to the choice of observer. 

Due to the above reasons, we need to answer the following questions in
order to clarify the state of the heat reservoir:

\begin{description}
\item[Q1.] What is a relativistic equilibrium state? 
Is the equilibrium state observer-dependent?
\item[Q2.] What is the equilibrium distribution for particles of the heat reservoir?
\item[Q3.] Is this distribution observer-dependent?
\end{description}

Fortunately the answers to these questions can be inferred from the studies on
relativistic kinetic theory. To answer Q1, let us infer that equilibrium states 
could be viewed as the final states of relaxation processes, and a system carrying out 
a relaxation process should not care about who is observing it. Therefore, the final state
of the relaxation process should not be affected by the choice of observer. Given 
an isolated system, there can only be one {\em intrinsic equilibrium state}, 
i.e. the state in detailed balance, which is characterized by several macroscopic 
features, including the absence of entropy production rate and vanishing collision integral
in the Boltzmann equation. 

From the point of view of the comoving observer, {\em Bob}, the equilibrium state
has one extra feature, i.e. the absence of transport flows. By definition, 
the proper velocity of Bob is identical to the proper 
velocity $U^\mu$ of an element of the heat reservoir viewed 
as a relativistic fluid. The same $U^\mu$ also appeared in the damping 
force term in the Langevin equation. 
According to \cite{liu2022covariant,hao2023gravitothermal}, the driving forces for the 
relativistic transports are the generalized gradients for the 
temperature and chemical potential, which are dependent on the proper velocity of 
the observer. For the comoving observer Bob, the generalized gradients for the 
temperature and chemical potential read
\begin{align}
\mathcal D_\nu T_{\rm B}=\nabla_\nu T_{\rm B} +T_{\rm B} U^\rho\nabla_\rho U_\nu=0,
\qquad
\mathcal D_\nu \mu_{\rm B}=\nabla_\nu \mu_{\rm B} 
+\mu_{\rm B} U^\rho\nabla_\rho U_\nu=0,
\label{gengr}
\end{align}
where $T_{\rm B}$ and $\mu_{\rm B}$ respectively are the temperature and 
chemical potential of the heat reservoir measured by Bob. One immediately sees that the 
ordinary gradients $\nabla_\nu T_{\rm B}$ and $\nabla_\nu \mu_{\rm B}$ are 
nonzero, unless Bob undergoes geodesic motion, i.e. $U^\rho\nabla_\rho U_\nu=0$. 
In the latter case, $T_{\rm B}$ and $\mu_{\rm B}$ becomes uniform, which is
fully consistent with the definition of equilibrium state in the non-relativistic 
thermodynamics.

The answer to Q2 is also provided by relativistic kinetic theory, and the explicit 
1PDF for the heat reservoir is given by the {\em J\"uttner-like distribution}\cite{cercignani2002relativistic} 
\begin{align}
\varphi_{\rm HR}(x,p) 
=\frac{\mathfrak g}{\re^{\alpha - B_\mu p^\mu}-\varsigma}
=\frac{\mathfrak g}{\re^{\alpha - U_\mu p^\mu/T_{\rm B}}-\varsigma}
=\frac{\mathfrak g}{\re^{(\varepsilon_{\rm B}-\mu_{\rm B})/T_{\rm B}}-\varsigma}
\label{Ju1}
\end{align}
provided that the background spacetime is stationary, 
where $\varsigma =0,\pm1$, $\mathfrak g$ denotes the quantum degeneracy, 
$\varepsilon_{\rm B}=- U_\mu p^\mu$ is the energy of the particle measured by Bob, 
$\mu_{\rm B}$ is the chemical potential of the heat reservoir, and 
$\alpha= - {\mu_{\rm B}}/{T_{\rm B}}$ is a constant in spacetime. 
In order that the distribution \eqref{Ju1} indeed describes a state 
in detailed balance, the vector field 
$B^\mu=U^\mu/T_{\rm B}$ is required to be timelike Killing, i.e.
\begin{align}
\nabla_{(\mu} B_{\nu)}=0.
\end{align}
The existence of a timelike Killing field implies that the underlying spacetime 
needs to be stationary.

We {\em assume that the heat reservoir is consisted of purely classical particles}.
In this case, the above 1PDF becomes the {\em standard  J{\" u}ttner distribution}
\begin{align}
\varphi_{\rm HR}(x,p) =\re^{-\alpha+ U_\mu p^\mu/T_{\rm B}}
=\re^{(\mu_{\rm B}-\varepsilon_{\rm B})/T_{\rm B}}.
\label{Ju2}
\end{align}
The 1PDF $\varphi_{\rm HR}(x,p)$ as presented in the form \eqref{Ju2} contains the 
proper velocity $U^\mu$ of Bob and the temperature $T_{\rm B}$ measured by Bob, thus 
it is explicitly dependent on the choice of observer. This answers Q3. 
It is remarkable that the distribution \eqref{Ju2} has the same form 
as the non-relativistic Boltzmann distribution. 
However, due to eq.\eqref{gengr}, the above distribution is in fact different 
from the Boltzmann distribution, because $T_{\rm B}$ and $\mu_{\rm B}$
are now non-uniform.

It is interesting to ask what the distribution \eqref{Ju2}
would look like from the point of view of Alice. To answer 
this question, let us remind that all measurements in curved spacetime must be 
made {\em on the spot}. Therefore, to consider the distributions of the same particles, 
Alice and Bob must appear at the same spacetime event, and their proper velocities 
can differ by at most a {\em local} Lorentz boost. Let $\gamma_{\rm AB}$ denotes the relative 
Lorentz factor between Alice and Bob. Then the proper velocity $U^\mu$ of Bob can 
be expressed as
\begin{align}
U^\mu = \gamma_{\rm AB} \left(Z^\mu +z^\mu\right), \quad
\gamma_{\rm AB}= -U_\mu Z^\mu,\quad
z^\mu Z_\mu=0,
\label{ZU}
\end{align}
The energy of the particle observed by Alice reads
$\varepsilon_{\rm A}=-Z_\mu p^\mu$.  Denoting the temperature and chemical potential of 
the heat reservoir measured by Alice as $T_{\rm A}$ and $\mu_{\rm A}$ respectively,  
we get, by inserting the eq.\eqref{ZU} into eq.\eqref{Ju2},
the following distribution,
\begin{align}
\varphi_{\rm HR}(x,p) 
&=\re^{-\alpha+\gamma_{\rm AB}(Z_\mu+z_\mu)p^\mu/T_{\rm B}}
=\re^{[\mu_{\rm B}+\gamma_{\rm AB}(Z_\mu+z_\mu)p^\mu]/T_{\rm B}}\nonumber\\
&=\re^{[(\gamma_{\rm AB})^{-1}\mu_{\rm B}+Z_\mu p^\mu+ z_\mu p^\mu]/T_{\rm A}}
= \re^{(\mu_{\rm A}- \varepsilon_{\rm A} + z_\mu p^\mu )/T_{\rm A}},
\end{align}
where the temperatures and the chemical potentials measured by both observers are related 
as \cite{hao2022relativistic}
\begin{align}
T_{\rm A}= (\gamma_{\rm AB})^{-1} T_{\rm B},\qquad
\mu_{\rm A}= (\gamma_{\rm AB})^{-1} \mu_{\rm B}.
\label{Tmutrans}
\end{align}

Now let us proceed with the second basic assumption for the Brownian motion un-altered,
hence the the probability distribution 
for the Brownian particle should approach the same form as the 1PDF 
for the heat reservoir after sufficient long time, i.e.
\begin{align}
\varphi(x,p)\to \varphi_{\rm HR}(x,p) 
=\re^{-\alpha+ U_\mu p^\mu/T_{\rm B}}
\quad {\rm as}\quad t\to\infty.
\label{juttner} 
\end{align}
The above assumption also implies that the long time
limit of the heat transfer rate $Q[\varphi]$ should approach zero, because 
for the J\"uttner distribution $\varphi$, we always have 
$\nabla_\mu T^{\mu\nu}[\varphi]=0$ and thus $Q[\varphi]=-Z_\nu \nabla_\mu T^{\mu\nu}[\varphi]
=0$. This result is independent of $Z_\mu$.
It is worth noting that the condition $Q[\varphi]=0$ does not necessarily imply 
$Z_\mu\mathcal I^\mu[\varphi]=0$. When the latter fails to vanish, 
it means that the Brownian particle is more likely to 
absorb heat from in some states and is more likely to transfer heat to 
the heat reservoir in some other states. Although the heat transfer 
between different micro states cancels out, this may lead to a deviation from 
detailed balance in the transition probabilities between different micro states, 
causing a change in the momentum distribution of the Brownian particle. 
Therefore, the {\em detailed} thermal equilibrium between the Brownian particle 
and the heat reservoir should be given by the stronger condition 
$Z_\mu\mathcal I^\mu[\varphi]=0$. Due to the arbitrariness in the choice of $Z^\mu$, 
this condition can be further reduced to $\mathcal I[\varphi]=0$, i.e.
\begin{align}\label{detailed-thermal-equilibrium}
\frac{1}{2}\mathcal{D}^{ij} \nabla^{(h)}_j\varphi
+\frac{\delta^{\fa\fb}}{2}\(\mathcal{R}^{i}{}_{\mathfrak{a}}  
\nabla^{(h)}_j\mathcal{R}^j{}_\mathfrak{b}\)\varphi
-\mathcal{F}^i_{\text{add}}\varphi-\mathcal{K}^{i\nu}U_\nu\varphi=0.
\end{align}
Inserting eq.\eqref{juttner} into eq.\eqref{detailed-thermal-equilibrium} we have
\begin{align}\label{extra-force}
\mathcal{F}^\mu_{\text{add}}
=\left[\frac{1}{2T_{\rm B}}\mathcal{D}^{\mu\nu}
-\mathcal{K}^{\mu\nu}\right]U_\nu
+\frac{\delta^{\fa\fb}}{2}\mathcal{R}^\mu{}_\mathfrak{a} 
\nabla^{(h)}_j\mathcal{R}^j{}_\mathfrak{b}.
\end{align}
In the low temperature limit, both $\mathcal{F}^\mu_{\text{add}}$ and 
$\displaystyle\frac{\delta^{\fa\fb}}{2}\mathcal{R}^\mu{}_\mathfrak{a} 
\nabla^{(h)}_j\mathcal{R}^j{}_\mathfrak{b}$ 
tends to vanish at least as $\mathcal{O}(T_{\rm B})$. Therefore we get
\begin{align}
\mathcal{D}^{\mu\nu}=2T_{\rm B}\mathcal{K}^{\mu\nu}+\mathcal O(T_{\rm B}^2),
\end{align}
where the extra $\mathcal O(T_{\rm B}^{2})$ term is dependent on the choice of damping model. 
When appropriate damping model is taken, e.g. like in \cite{klimontovich1994nonlinear},
this term can be removed completely, yielding
\begin{align}\label{equation-einstein-relation}
\mathcal{D}^{\mu\nu}=2T_{\rm B} \mathcal{K}^{\mu\nu}.
\end{align}
This relation is the general relativistic analogue of the celebrated Einstein relation.

As a simple intuitive example case, let us consider the isotropic damping model 
in which the diffusion tensor and tensorial damping coefficients are given as
\begin{align}
\mathcal{K}^{\mu\nu}=\kappa \Delta^{\mu\nu}(p),\qquad
\mathcal{D}^{\mu\nu}=D \Delta^{\mu\nu}(p),
\end{align}
where $k$ and $D$ are both scalar functions, i.e. the scalar friction coefficient 
and diffusion coefficient respectively. Then the Einstein relation will 
degenerate into
\begin{align}
D=2\kappa T_{\rm B},
\end{align}
which is formally identical to that arises from non-relativistic linear response theory,
except that $T_{\rm B}$ now could have a nonvanishing ordinary gradient 
because of eq.\eqref{gengr}. This result suggests that linear response theory 
should still hold in the relativistic context, at least locally.

When the Einstein relation \eqref{equation-einstein-relation} holds precisely, 
we have
\begin{align}
\mathcal{F}^{\mu}_{\text{add}}
=\frac{\delta^{\fa\fb}}{2}\mathcal{R}^{\mu}_{\mathfrak{a}} 
\nabla^{(h)}_{j}\mathcal{R}^{j}_{\mathfrak{b}}.
\label{Fadd}
\end{align}
This result has already been adopted in \cite{cai2023relativistic} while constructing 
the general covariant Langevin equations. Inserting eq.\eqref{Fadd} into the 
definition of $\mathcal I[\varphi]$ yields
\begin{align}
\mathcal I[\varphi]=\left[ \frac{1}{2}\mathcal{D}^{ij} \nabla^{(h)}_j\varphi
-\mathcal{K}^{i\nu}U_\nu\varphi  \right]\frac{\partial }{\partial \breve p^i}.
\end{align}
This formula provides a physical image of how the Brownian particle reaches 
equilibrium after long time of relaxation. The damping force causes a heat transfer 
from the Brownian particle to the heat reservoir, while the stochastic force 
causes a heat transfer form the heat reservoir to the Brownian particle. 
After long time of relaxation, the damping and stochastic forces balances 
each other in the statistical sense.

As a final remark, let us mention that, due to the transformation rule \eqref{Tmutrans}, 
the Einstein relation rewritten in terms of the temperature measured 
by Alice should read
\[
\mathcal{D}^{\mu\nu}=2\gamma_{\rm AB} T_{\rm A} \mathcal{K}^{\mu\nu}.
\]

\section{Conclusion}\label{conclusion}

The major results of the present work can be summarized as follows.

1) The general covariant FPEs associated with both versions
of the general relativistic Langevin equation proposed in Ref.\cite{cai2023relativistic}
are presented, both give rise to the same reduced FPE 
obeyed by the 1PDF $\varphi(x,p)$ for the Brownian particle. The relationship between 
different distribution functions is clarified. 

2) Several important macroscopic quantities and the quantitative relationships between them 
are obtained with the aid of the 1PDF which obeys the reduced FPE. 
These quantities and relationships reveal a close connection between the approaches 
of stochastic mechanics and relativistic kinetic theory.

3) The meaning of the relativistic equilibrium state of the heat reservoir 
is properly addressed, and, by assuming that the long time relaxation result for the 1PDF
of the Brownian particle should be identical to the 1PDF of the heat reservoir, 
we derive a general covariant version of the Einstein relation. 

These results resolve several common confusions which exist in the literature. Moreover, 
we hope to use these results as the starting point for further exploring some 
important subjects in relativistic macroscopic systems, e.g. the origin of 
irreversibility in relativistic systems, the area law of near horizon entropies, etc. 
More on these topics will come about later.

\section*{Appendex A: Diffusion operator approach to the FPE}
\label{diffusion-operator}

In order to derive the FPE from a stochastic differential 
equation (SDE), we need to use Ito's lemma to calculate the differential of an 
arbitrary scalar function, and perform integration by parts twice. When the SDE 
is defined on a manifold, this procedure can be very complicated. 

There is a simpler approach, i.e. the diffusion operator approach \cite{bakry2014analysis}, 
for obtaining the FPE on a manifold. Here we give a brief 
review of this alternative method.

The Ito type SDE on Riemannian manifold or Pseudo-Riemannian manifold $(M,g)$ 
can be written as
\begin{align}
\D\tilde X^\mu_t=F^\mu \D t+ C^\mu{}_{\fa} \circ_I \D\tilde w_t^{\fa}.
\label{SDE1}
\end{align}
Let $h$ be an arbitrary scalar field on $M$, then the time differential of 
$\tilde h_t:=h(\tilde X_t)$ can be derived by Ito's lemma:
\begin{align}
\D \tilde h_t=\left[ \frac{\partial  h}{\partial x^{\mu }} F^{\mu} 
+\frac{\delta^{\fa\fb}}{2} \frac{\partial^2 h}{\partial x^{\mu } 
\partial x^{\nu}} C^{\mu}{}_{\fa} C^{\nu}{}_{\fb} \right]\D t
+ \frac{\partial h}{\partial x^\mu}C\ind{\mu}{\fa}\circ_I \D\tilde w_t^\fa.
\end{align}
Therefore, the expectation of $\D \tilde h_t$ is
\begin{align}
\langle \D \tilde h_t\rangle
=\left\langle \frac{\partial  h}{\partial x^{\mu }} F^{\mu} 
+\frac{\delta^{\fa\fb}}{2} \frac{\partial^2 h}{\partial x^{\mu } 
\partial x^{\nu}} C^{\mu}{}_{\fa} C^{\nu}{}_{\fb} \right\rangle \D t .
\end{align}
This means $\langle\tilde h_t\rangle$ is differentiable with respect to time 
in spite of the fact that $\tilde h_t$ isn't differentiable. Defining the 
diffusion operator as
\begin{align}\label{diffusion_operation_Ito}
\mathbf{A}=F^\mu \frac{\partial}{\partial x^\mu}
+\frac{\delta^{\fa\fb}}{2}C^\mu{}_{\fa} C^\nu{}_{\fb}\frac{\partial^2}
{\partial x^\mu \partial x^\nu},
\end{align}
the derivative of $\langle\tilde h_t\rangle$ can be written as
\begin{align}
\frac{\D}{\D t}\langle\tilde h_t\rangle=\langle \mathbf{A} \tilde h_t \rangle.
\end{align}

Let $\Phi_t(x):=\Pr[\tilde X_t=x]$ be a PDF associated with the invariant 
volume element $\sqrt{g}\D^n x$ of $M$, above equation actually means
\begin{align}
\frac{\D}{\D t}\int_M h \Phi_t \sqrt{g} \D^n x
=\int_M h \partial_t \Phi_t \sqrt{g} \D^n x
=\int_M \Phi_t \mathbf{A} h \sqrt{g} \D^n x
=\int_M (\mathbf{A}^* \Phi_t) h  \sqrt{g} \D^n x,
\end{align}
where $\mathbf{A}^*$ is the adjoint of $\mathbf{A}$. 
Since $h$ is arbitrary, the above equation implies
\begin{align}
\partial_t \Phi_t=\mathbf{A}^* \Phi_t,
\label{diffuFP}
\end{align}
which is the FPE associated with the SDE \eqref{SDE1}. 

There are four rules for computing the adjoint operator:
\begin{enumerate}
\item $(\mathbf{A}+\mathbf{B})^*=\mathbf{A}^*+\mathbf{B}^*$.
\item $(\mathbf{AB})^*=\mathbf{B}^*\mathbf{A}^*$.
\item $\displaystyle \(\frac{\partial}{\partial x^\mu}\)^*
=-\frac{1}{\sqrt{g}}\frac{\partial}{\partial x^\mu}\sqrt{g}$, where the right hand side needs
to be understood as a right associative operator.
\item $(F^\mu)^*=F^\mu$.
\end{enumerate}
Using these rules, the adjoint of the diffusion operator \eqref{diffusion_operation_Ito}
is evaluated to be
\begin{align}
\mathbf{A}^* &=-\frac{1}{\sqrt{g}}\frac{\partial}{\partial x^\mu}\sqrt{g} F^\mu 
+\frac{\delta^{\fa\fb}}{2}\frac{1}{\sqrt{g}}\frac{\partial^2}{\partial x^\mu \partial x^\nu} \sqrt{g} C^\mu{}_{\fa} C^\nu{}_{\fb}.
\end{align}

Since the Stratonovich type SDE 
\begin{align}
\D \tilde X^\mu=F^\mu \D t+ C^\mu{}_{\fa} \circ_S \D \tilde w^{\fa} 
\label{StratonovichSDE}
\end{align}
is equivalent to the Ito type SDE
\begin{align}
\D \tilde X^{\mu}=\(F^{\mu}+\frac{\delta^{\fa\fb}}{2}C^{\nu}{}_{\fa}
\frac{\partial}{\partial x^{\nu}} C^{\mu}{}_{\fb}\) \D t
+ C^{\mu}{}_{\fa} \circ_I \D \tilde w^{\fa},
\end{align}
the corresponding diffusion operation reads
\begin{align}
\mathbf{A}&=\(F^\mu+\frac{\delta^{\fa\fb}}{2}C^\nu{}_{\fa}
\frac{\partial}{\partial x^\nu} C^\mu{}_{\fb}\)\frac{\partial }{\partial x^\mu}
+\frac{\delta^{\fa\fb}}{2}C^\nu{}_{\fa} C^\mu{}_{\fb} 
\frac{\partial^2}{\partial x^\mu \partial x^\nu}\nonumber\\
&=F^\mu \frac{\partial}{\partial x^\mu}
+\frac{\delta^{\fa\fb}}{2}C^\nu{}_{\fa}\frac{\partial}{\partial x^\nu}
 C^\mu{}_{\fb} \frac{\partial}{\partial x^\mu}.
\end{align}
Introducing the vector fields
\begin{align}
L_0=F^\mu\frac{\partial}{\partial x^\mu}\qquad L_{\fa}
=C^\mu{}_{\fa}\frac{\partial}{\partial x^\mu},
\end{align}
the diffusion operation can be written as simpler form
\begin{align}
\mathbf{A}=\frac{\delta^{\fa\fb}}{2}L_{\fa} L_{\fb} +L_0.
\end{align}
It is easy to see that $L_0$ provides the drift term of FPE 
and $L_{\fa}$ provides the diffusion term. Notice that the adjoint of the coordinate 
derivative operator looks like the covariant divergence operator when acting on 
a vector field. Therefore, the action of the adjoint of $\mathbf{A}$ on the 
PDF becomes
\begin{align}
\mathbf{A}^*\Phi_t&=\frac{\delta^{\fa\fb}}{2}L^*_{\fa} L^*_{\fb}\Phi_t +L^*_0\Phi_t\notag\\
&=\frac{\delta^{\fa\fb}}{2}\nabla_\mu( C^\mu{}_{\fa}(\nabla_\nu C^\nu{}_{\fb}\Phi_t))
-\nabla_\mu (F^\mu\Phi_t).
\end{align}
Inserting this result into eq.\eqref{diffuFP} gives rise to the Fokker-Planck 
equation associated with the Stratonovich type SDE \eqref{StratonovichSDE}.

\section*{Acknowledgement}

This work is supported by the National Natural Science Foundation of China under the grant
No. 12275138.

\section*{Data Availability Statement} 

This research has no associated data. 

\section*{Declaration of competing interest}

The authors declare no competing interest.

\providecommand{\href}[2]{#2}\begingroup
\footnotesize\itemsep=0pt
\providecommand{\eprint}[2][]{\href{http://arxiv.org/abs/#2}{arXiv:#2}}

\end{document}